\pdfoutput=1
\documentclass[11pt,twoside,a4paper,cmspaper,final,collab]{cms-tdr}
\def\svnVersion{9e6b80d}\def\svnDate{2023/10/23}\def\cmsCernNoTag{CERN-EP-2023-052}\def\cmsCernDate{\today}\def\cmsMessage{Published in the Journal of High Energy Physics as \href{http://dx.doi.org/10.1007/JHEP10(2023)115}{\doi{10.1007/JHEP10(2023)115}.}}
\begin{document}\cmsNoteHeader{HIN-21-008}

\newcommand{\upsifam}{\ensuremath{\Upsilon}(1S, 2S, 3S)\xspace}
\newcommand{\vtwo}{\ensuremath{v_{2}}\xspace}
\newcommand{\vn}{\ensuremath{v_{n}}\xspace}
\newcommand{\vni}{\ensuremath{v^i_{n}}\xspace}
\newcommand{\vthree}{\ensuremath{v_{3}}\xspace}
\newcommand{\relphantom}[1]{\mathrel{\phantom{#1}}}
\newcommand{\sqrts}{\ensuremath{\sqrt{s}}\xspace}
\newcommand{\mumu}{\ensuremath{\PGmp\PGmm}\xspace}

\newcommand{\Npart}{\ensuremath{\langle\mathrm{N}_\text{part}\rangle}\xspace}

\newcommand{\Lxyz}{\ensuremath{L_{xyz}}\xspace}
\newcommand{\ctau}{\ensuremath{\ell_{\PJGy}}\xspace}
\newcommand{\ctaus}{\ensuremath{\ell_{\PJGy,\PGyP{2S}}}\xspace}
\newcommand{\ctautwos}{\ensuremath{\ell_{\PGyP{2S}}}\xspace}
\newcommand{\ljpsi}{\ensuremath{\ell_{\PJGy}}\xspace}
\newcommand{\lpsi}{\ensuremath{\ell_{\PGy}}\xspace}
\newcommand{\ctaui}{\ensuremath{\ell_{\PJGy,i}}\xspace}
\newcommand{\ctauxy}{\ensuremath{\ell_{\PJGy}^\text{2D}}\xspace}
\newcommand{\ctauxyz}{\ensuremath{\ell_{\PJGy}^\text{3D}}\xspace}

\newcommand{\NSig}{\mathcal{N}_\text{sig}\xspace}
\newcommand{\NBkg}{\mathcal{N}_\text{bkgd}\xspace}
\newcommand{\RAA}{\ensuremath{R_\mathrm{AA}}\xspace}
\newcommand{\PbPb}{\ensuremath{\text{PbPb}}\xspace}
\newlength\cmsTabSkip\setlength{\cmsTabSkip}{1ex}

\providecommand{\cmsLeft}{left\xspace}
\providecommand{\cmsRight}{right\xspace}
\providecommand{\appLabel}{Appendix}

\cmsNoteHeader{HIN-21-008} 
\title{Measurements of the azimuthal anisotropy of prompt and nonprompt charmonia in \texorpdfstring{\PbPb}{PbPb} collisions at \texorpdfstring{$\sqrtsNN = 5.02\TeV$}{sqrt(s[NN]) = 5.02 TeV}}

\date{\today}

\abstract{The second-order (\vtwo) and third-order (\vthree) Fourier coefficients describing the azimuthal anisotropy of prompt and nonprompt (from \Pb{}-hadron decays) \PJGy, as well as prompt \PGyP{2S} mesons are measured in lead-lead collisions at a center-of-mass energy per nucleon pair of $\sqrtsNN = 5.02\TeV$. The analysis uses a data set corresponding to an integrated luminosity of 1.61\nbinv recorded with the CMS detector. The \PJGy and \PGyP{2S} mesons are reconstructed using their dimuon decay channel. The \vtwo and \vthree coefficients are extracted using the scalar product method and studied as functions of meson transverse momentum and collision centrality. The measured \vtwo values for prompt \PJGy mesons are found to be larger than those for nonprompt \PJGy mesons. The prompt \PJGy \vtwo values at high \pt are found to be underpredicted by a model incorporating only parton energy loss effects in a quark-gluon plasma medium. Prompt and nonprompt \PJGy meson \vthree and prompt \PGyP{2S} \vtwo and \vthree values are also reported for the first time, providing new information about heavy quark interactions in the hot and dense medium created in heavy ion collisions.}

\hypersetup{
pdfauthor={CMS Collaboration},
pdftitle={Measurements of the azimuthal anisotropy of charmonia in PbPb collisions at sqrtsNN = 5.02 TeV},
pdfsubject={CMS},
pdfkeywords={CMS, PbPb, quarkoninum, heavy quark, charmonium, flow, v2, v3}}

\maketitle 

\section{Introduction}
In high energy heavy ion collisions, a very strongly-interacting medium of deconfined quarks and gluons, known as the quark-gluon plasma (QGP), is created~\cite{Busza:2018rrf}.
Quarkonia, bound states of a heavy quark and its antiquark, e.g., charmonia (\PJGy, \PGyP{2S}), and bottomonia (\upsifam), are useful probes to study the properties of the QGP.
Since heavy quarks are predominantly produced at the early stages of collisions by hard partonic scattering processes, they experience the whole space-time evolution of the medium.
Models including static color screening~\cite{Matsui:1986dk,Digal:2001ue} and gluo-dissociation effects~\cite{laine:2007,Brambilla:2008cx,Brambilla:2010vq} inside a QGP medium predict a suppression of quarkonium yields in nucleus-nucleus (AA) collisions compared to proton-proton ($\Pp\Pp$) collisions~\cite{Rothkopf:2019ipj}.
On the other hand, quarkonia may also be created by recombination of quarks and antiquarks which were initially uncorrelated, thereby resulting in an enhancement of the measured yields~\cite{blaizot}.

Experimentally, the modification of particle yields in heavy ion collisions is quantified by the nuclear modification factor, defined as the ratio of yields in AA collisions to those in $\Pp\Pp$ collisions scaled by the estimated number of binary nucleon-nucleon (NN) collisions.
Results from the CERN SPS, BNL RHIC, and CERN LHC experiments show significant suppression of \PJGy mesons in AA collisions~\cite{NA38:1994yau,NA50:2004sgj,NA60:2007ewx,Capella:1988ha,PHENIX:2006gsi,PHENIX:2011img,STAR:2013eve,STAR:2016utm,Abelev:2012rv,Adam:2016rdg,CMS:2017uuv}.
At the LHC, the ALICE Collaboration~\cite{Abelev:2012rv,Adam:2016rdg} observed a weaker suppression of \PJGy mesons in the low transverse momentum (\pt) region below 3~\GeVc than was seen by experiments at the SPS and RHIC.
This result has been interpreted as a sign of enhanced recombination processes at LHC energies because of the larger production cross section of charm quarks~\cite{Abelev:2012rv,Adam:2016rdg,Zhao:2011cv,Du:2015wha}.
These measurements have provided information to constrain theoretical models that incorporate different in-medium effects on charmonium states in order to describe thermal characteristics of the QGP.

Another effective way to probe the dynamics of the QGP is the study of the azimuthal ($\phi$) distribution of particles produced in heavy ion collisions.
Azimuthal anisotropies of the particles can be characterized by Fourier coefficients (\vn) of the $\phi$ distribution with respect to the event plane, which corresponds to the direction of maximal particle density~\cite{Voloshin:1994mz}.
The presence of non-zero Fourier coefficients is referred to as collective flow by analogy to hydrodynamic models.
In particular, the second-order (\vtwo) and the third-order (\vthree) components are predominantly sensitive to initial collision geometry and event-by-event fluctuations, respectively~\cite{Bzdak:2013zma}.
Because of the initial anisotropic shape of the QGP in the transverse plane, quarkonia produced by initial hard-scattering processes experience different path lengths while traveling through the medium.
Larger energy loss, and therefore larger suppression, can occur for quarkonia moving in directions corresponding to larger average path lengths.
On the other hand, quarkonia created via recombination can have an asymmetric $\phi$ distribution as a result of asymmetries in quark densities~\cite{Liu:2010ef,He:2021zej}.

In LHC experiments, sizable \vtwo values have been observed for \PJGy mesons in lead-lead (\PbPb) collisions at center-of-mass energies per nucleon pair $\sqrtsNN$ = 2.76 and 5.02\TeV~\cite{Abbas:2013jl,ATLAS:Jpsiv2,Khachatryan:2017sz,Acharya:2020jf}, indicating collective flow behavior of charm quarks.
However, these results have limitations that prevent drawing firm conclusions on the origin of the \vtwo for \PJGy mesons, given the large measurement uncertainties and the lack of separation between prompt and nonprompt \PJGy mesons.
Measurements of nonprompt \PJGy mesons, which originate from bottom hadron decays, provide useful information on the propagation of bottom quarks in the QGP.

The CMS Collaboration has shown that in PbPb collisions the suppression of prompt \PJGy (of \pt above 6.5~\GeVc) within jets depends upon the fragmentation~\cite{CMS:2021puf}.
Theoretical calculations suggest that jet quenching may drive the modulation of the $\phi$ distribution of \PJGy at high \pt~\cite{Zhang:2022rby}.
These studies motivate the measurement of the azimuthal anisotropy for prompt \PJGy mesons at higher \pt to investigate the path length dependence of jet quenching~\cite{Sirunyan:2018aa,CMS:2012tqw}.

Azimuthal correlations for \PGyP{2S} mesons are also a subject of interest.
Prompt \PGyP{2S} mesons have been found to be significantly more suppressed than prompt \PJGy mesons in \PbPb collisions~\cite{CMS:2017uuv,ALICE:2022jeh}.
Measurements of \vtwo and \vthree values for prompt \PGyP{2S} mesons are expected to provide constraints on in-medium effects for charmonium states with different binding energies in heavy ion collisions~\cite{Du:2015wha,Chen:2016mhl}.

This paper reports measurements of \vn values for prompt and nonprompt \PJGy mesons, as well as prompt \PGyP{2S} meson, using \PbPb data collected with the CMS detector at $\sqrtsNN$ = 5.02\TeV in 2018.
The prompt and nonprompt \PJGy \vthree, and prompt \PGyP{2S} \vn values are measured for the first time.
The charmonium states are identified via their dimuon decay channel.
The data set corresponds to an integrated luminosity of 1.61\nbinv~\cite{CMS-PAS-LUM-18-001,CMS:2021xjt}, allowing the extraction of \vn values as functions of meson transverse momentum and \PbPb event centrality (i.e., the degree of overlap of the two Pb ions).
The numerical values of the results for this analysis are tabulated in the HEPDATA record~\cite{HEPData}.

\section{The CMS detector}
\label{sec:cmsdetector}
The central feature of the CMS apparatus is a superconducting solenoid of 6\unit{m} internal diameter, providing a magnetic field of 3.8\unit{T}.
Within the solenoid volume are a silicon pixel and strip tracker, a lead tungstate crystal electromagnetic calorimeter (ECAL), and a brass and scintillator hadron calorimeter (HCAL), each composed of a barrel and two endcap sections.
Forward calorimeters extend the pseudorapidity ($\eta$) coverage provided by the barrel and endcap detectors.
The forward hadron (HF) calorimeter uses steel as an absorber and quartz fibers as the sensitive material.
The two halves of the HF are located 11.2\unit{m} from the interaction region, one on each end, and together they provide coverage in the range $3.0 < \abs{\eta} < 5.2$.
The HF calorimeters are composed of individual detector elements, or towers, having $\Delta \eta \times \Delta \phi = 0.175 \times 0.175$.
Centrality is determined by the total transverse energy deposited in both of the HF calorimeters, and is defined as the fraction of the total hadronic inelastic nucleus-nucleus cross section, with 0\% representing the largest overlap of the two colliding Pb nuclei~\cite{LoizidesImprovedMC}.

Muons are measured in the range $\abs{\eta} < 2.4$, with detection planes made using three technologies: drift tubes, cathode strip chambers, and resistive plate chambers.
Matching muons to tracks measured in the silicon tracker results in a relative transverse momentum resolution, for muons with \pt up to 100~\GeVc, of 1\% in the barrel and 3\% in the endcaps.
The \pt resolution in the barrel is better than 7\% for muons with \pt up to 1~\TeVc~\cite{CMS:2018rym}.
A more detailed description of the CMS detector, together with a definition of the coordinate system used and the relevant kinematic variables, can be found in Ref.~\cite{CMS:2008xjf}.

\section{Event selection}
\label{sec:selection}
Events of interest are selected using a two-tiered trigger system.
The first level (Level-1), composed of custom hardware processors, uses information from the calorimeters and muon detectors to select events at a rate of around 100\unit{kHz} within a fixed latency of about 4\mus~\cite{CMS:2020cmk}.
The second level, known as the high-level trigger (HLT), consists of a farm of processors running a version of the full event reconstruction software optimized for fast processing, and reduces the event rate to around 1\unit{kHz} before data storage~\cite{Khachatryan:2016bia}.
The trigger for this analysis was designed to have several stages.
First, at least two muon candidates should be reconstructed at Level-1.
At the HLT, the events are required to contain at least one Level-2 (L2) muon and another Level-3 (L3) muon, with requirements on their invariant mass being within 1--5 \GeVcc.
L2 muons are identified by matching the tracks to the hits in the outer muon spectrometer.
For L3 muons, full tracks from the L2 muon tracks and the inner tracker information are used in the reconstruction, in addition requiring at least ten high-quality hits in the inner tracker~\cite{Khachatryan:2016bia}.

To reject beam-related background processes, events are required to have at least one reconstructed primary vertex, and the clusters in the silicon pixel detector are required to be compatible with the vertex position. 
There must be at least two towers that contain energy deposits above 4 \GeV in each of the HF detectors.
The primary vertex, which is reconstructed from two or more tracks, is required to be located within 15\unit{cm} of the central point of the detector along the beam axis.
In addition, the reconstructed muons are selected using a set of offline muon identification criteria~\cite{CMS:2021puf}.
To ensure the reconstruction efficiency of single muons is larger than 10\%, muons within the following kinematic domains are selected:
\begin{equation}
\begin{aligned}
  &\pt^{\mu} > 3.5 \GeVc & \text{ for   } & \abs{\eta^{\mu}}<1.2, \\
  &\pt^{\mu} > (5.47 - 1.89\abs{\eta^{\mu}}) \GeVc & \text{ for   } & 1.2<\abs{\eta^{\mu}}<2.1,~\mathrm{and} \\
  &\pt^{\mu} > 1.5 \GeVc & \text{ for   } & 2.1<\abs{\eta^{\mu}}<2.4. \\
\end{aligned}
\label{eqn:acccut}
\end{equation}

\section{Acceptance and efficiency corrections}
\label{sec:corrections}
Correction factors are applied to all results to account for the detector acceptance, as well as for the trigger, reconstruction, and selection efficiencies for \mumu pairs.
The correction factors are derived from simulated prompt and nonprompt \PJGy and \PGyP{2S} meson samples which are then embedded in simulated PbPb collisions. 
Corrections are evaluated in the same bins of \pt, centrality, and rapidity ($y$) used in the \vn analyses.
The meson samples are generated with \PYTHIA 8.212~\cite{Sjostrand2015} using the CP5 underlying event tune~\cite{Sirunyan:2019dfx}, while the decay of bottom hadrons is simulated with \EVTGEN 1.3.0~\cite{Lange:2001uf}, which provides a better description of the kinematic distribution of nonprompt charmonia.
These meson events are then embedded into simulated PbPb collision events generated using \HYDJET 1.9~\cite{Gale:2013da}, and propagated through the CMS detector with the \GEANTfour package~\cite{Agostinelli:2002hh}.
The \pt distributions for simulated \PJGy and \PGyP{2S} mesons are compared to those in data with fine \pt intervals, and the ratios of data over the simulation are used to reweight the simulation to better describe the data as a function of \pt.
This weighting procedure accounts for possible mismodeling of the \PJGy and \PGyP{2S} meson kinematics.
Moreover, the reconstructed charmonia in data include feed-down contributions from heavier quarkonium state decays.
These contributions are not explicitly considered in the simulation, but their effect on the kinematic distribution of the simulated charmonia is taken into account by the \pt reweighting procedure.

The acceptance in a given analysis bin is defined as the fraction of generated \PJGy and \PGyP{2S} mesons in that bin that decay into two muons passing the kinematic requirements defined in Eq.~(\ref{eqn:acccut}), and reflects the geometrical coverage of the CMS detector.
The value of the acceptance increases with dimuon \pt and varies from 0.05--0.76.

The efficiency in a given analysis bin is defined as the ratio of the number of reconstructed \PJGy and \PGyP{2S} mesons from which both muons pass the selections described in Section~\ref{sec:selection} and the number of generated \PJGy mesons from which both muons pass the kinematic requirements defined in Eq.~(\ref{eqn:acccut}).
Individual components of the efficiency (trigger, track reconstruction, and muon identification) are measured using single muons from \PJGy meson decays in both collision data and simulated samples, using a ''tag-and-probe'' (T\&P) technique~\cite{Khachatryan:2011aa,Chatrchyan:2012np}.
The ratio in the efficiency between the data and simulated samples is used as a correction factor for the efficiency extracted from the simulation.
The reciprocals of the acceptance and efficiency corretion factors are used as event-by-event weights when the distributions of invariant mass and average \vn are computed.

\section{Signal extraction}
\label{sec:SinalExt}
The signal candidates for \PJGy (\PGyP{2S}) mesons are extracted in the kinematic range of $6.5 < \pt < 50$ \GeVc, $\abs{y} < 2.4$, and centrality 0--90\% (0--60\%).
The results are reported as functions of \pt and \Npart, where \Npart is the average number of participating nucleons in a given centrality interval~\cite{LoizidesImprovedMC}.
The centrality range for \pt-dependent results is limited to 10--60\% to ensure a large anisotropy of the QGP. 
The most central collisions (0--10\%) are excluded due to the small eccentricity in the initial geometry of the medium, while the most peripheral collisions (60--90\%) are excluded because of the poor resolution in extracting the $v_n$ coefficients. 
In addition, the lower \pt region is extended down to 3 \GeVc (4\GeVc) for \PJGy (\PGyP{2S}) in the forward rapidity region.
Details of the kinematic selection and \Npart values for each centrality interval can be found in Tables~\ref{tab:KinematicRegion} and \ref{tab:CentralityNpart} in \appLabel~\ref{app:KinematicRegion}.
The mass ranges of 2.6--3.5 and 3.3--4.1 \GeVcc are used to study the \PJGy and \PGyP{2S} mesons, respectively.

\begin{figure}[hbt]
  \centering
  \includegraphics[width=0.48\textwidth]{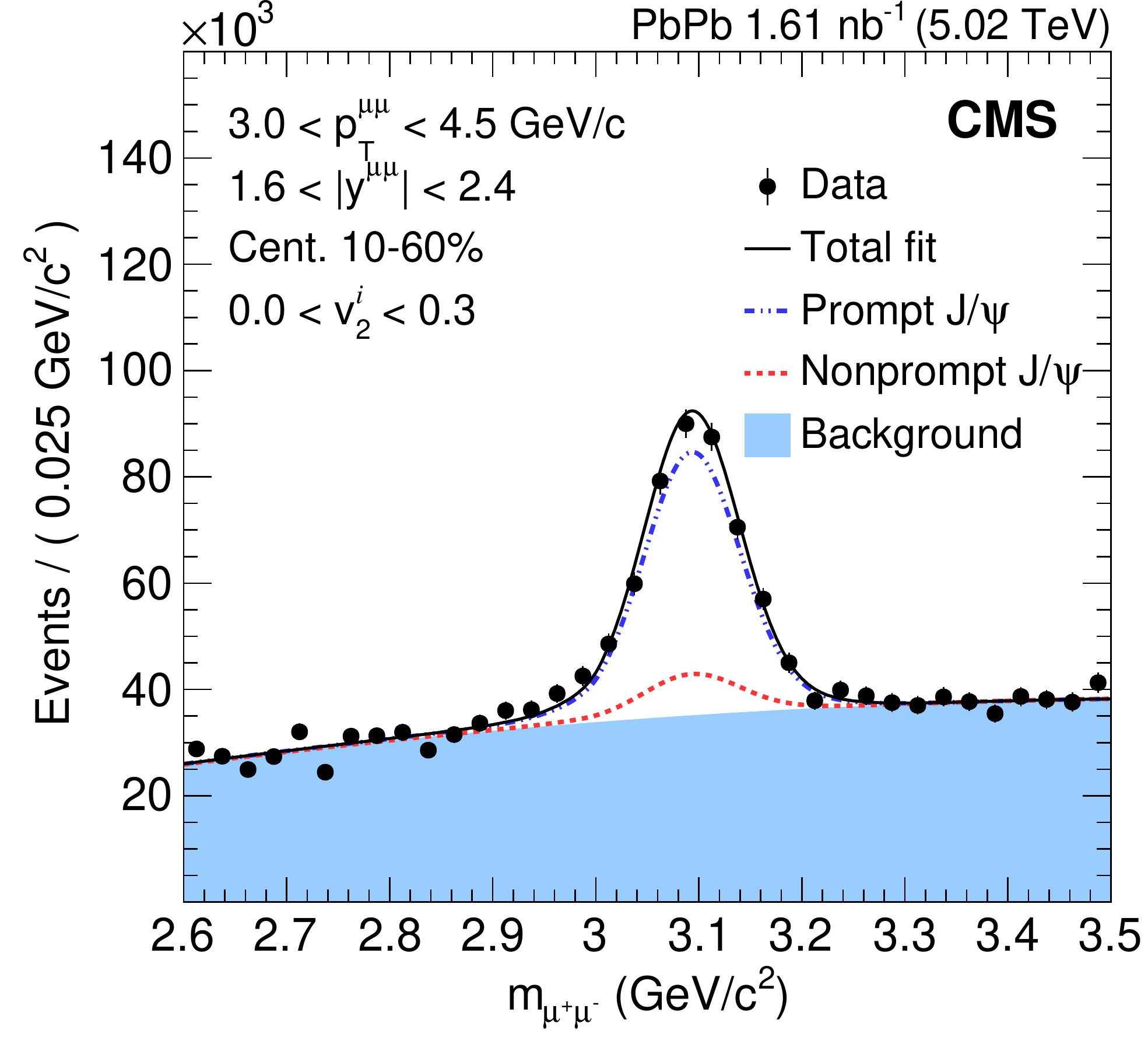}
  \includegraphics[width=0.48\textwidth]{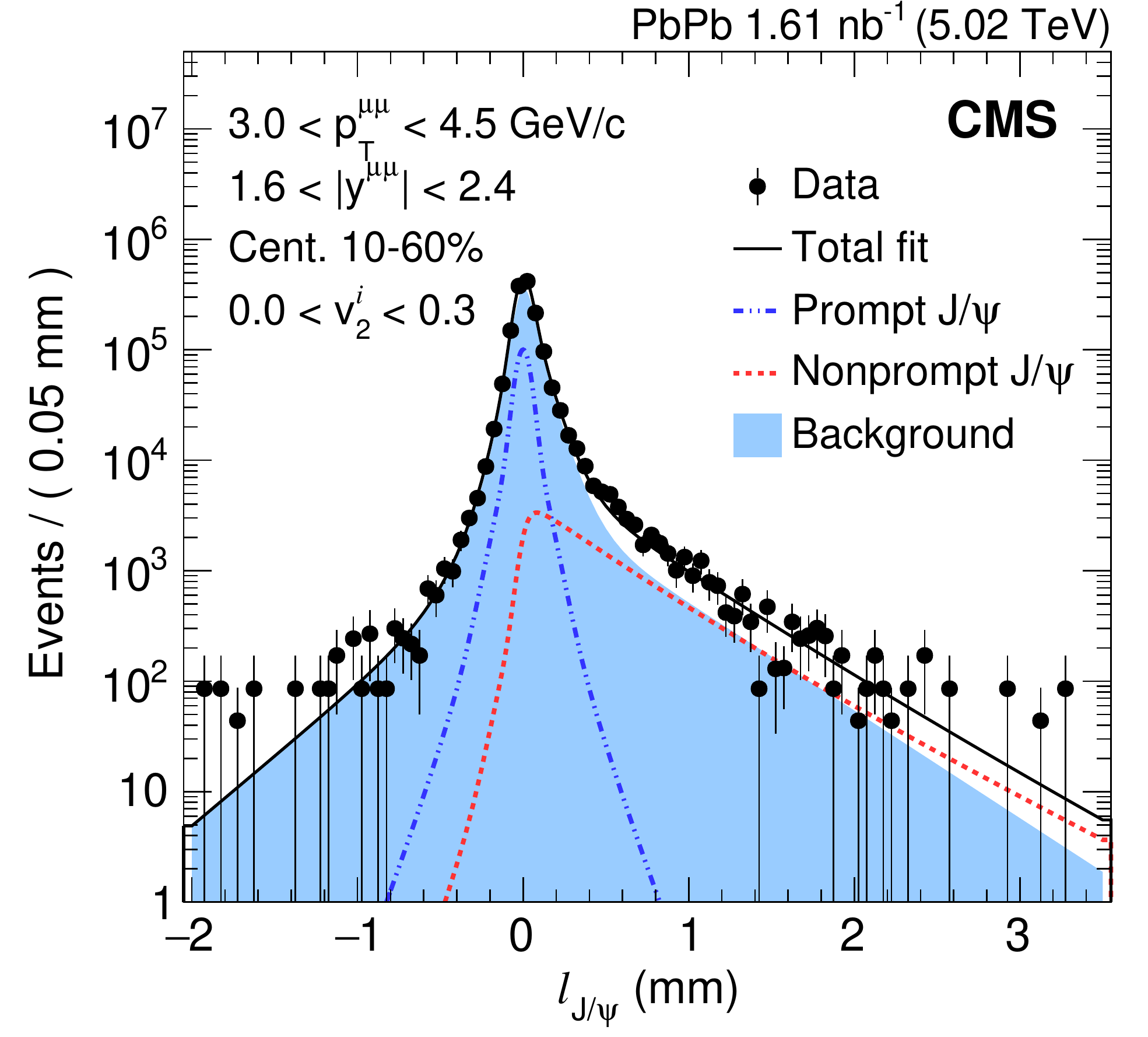}
  \caption{Invariant mass (\cmsLeft) and $\ell_{\PJGy}$ (\cmsRight) distributions for the bin of $3.0 < \pt < 4.5$ \GeVc and $0.0 < \vtwo < 0.3$ in centrality 10--60\%. The solid lines represent the total fit, while the dashed and dash--dotted lines represent the prompt and nonprompt components, respectively. The background contributions are shown by the filled blue histograms.}
\label{fig:2DFits}
\end{figure}

The separation of the prompt and nonprompt charmonium production components relies on the measurement of a secondary $\PGmp\PGmm$ vertex displaced from the primary collision vertex.
The pseudo-proper decay length~\cite{Buskulic:1993aa} is defined as $\ctau =  \Lxyz \: m_{\PJGy} \:c / |p_{\mu\mu}|$, where $\Lxyz$ is the distance between the primary and dimuon vertices, $m_{\PJGy}$ is the world average value~\cite{Zyla:2020zbs} of the \PJGy meson mass (assumed for all dimuon candidates), and $p_{\mu\mu}$ is the dimuon momentum.
To extract prompt and nonprompt \PJGy meson yields, the invariant mass spectrum of $\PGmp\PGmm$ pairs and their \ctau distribution are fitted using a two-dimensional (2D) extended unbinned maximum likelihood fit.
The parameters of different components of the 2D probability density function (pdf) are obtained through fits to the invariant mass and the \ctau distributions, which are further used in the final 2D fits. 
These fits are performed for each \vn interval in the measured kinematic region.
The evaluation of \vn coefficients is detailed in Section~\ref{sec:v2ext}.
Figure~\ref{fig:2DFits} shows the dimuon invariant mass and \ctau projection of the 2D fit in the ranges $3.0 < \pt < 4.5$ \GeVc and $0.0 < \vtwo < 0.3$ for the 10--60\% centrality interval.

Sums of two Crystal Ball (CB) functions~\cite{SLAC-R-236}, with different widths but common mean and tail parameters, are used to extract the signal yield values from the invariant mass distribution.
Two separate sums of CB functions are used to describe the prompt and nonprompt contributions to the \PJGy signal.
The tail parameters, as well as relative contributions of the two CB functions and the ratio of their widths, are fixed to the values obtained from simulation.
The background component for the invariant mass distribution is described by a sum of Chebyshev polynomial functions up to order $N$, where $N$ is determined by performing the log-likelihood ratio (LLR) test~\cite{Cowan:2010js}.
This procedure is done for each analysis bin, while the tail and width ratio parameters are kept fixed.
The polynomial order is chosen as the minimum so that increasing it does not significantly improve the quality of the fit.
In most analysis bins, the first-order Chebyshev polynomial function is used as the background function.

Prompt and nonprompt signals and background components of \ctau distributions are parameterized using collision data and simulated samples.
The {\it sPlot\/} technique~\cite{sPlot} is used to obtain \ctau distributions of \PJGy signal and background from data, which have different invariant mass distributions.
The correlation between the mass and \ctau is found to be very small, such that its impact on the final results is negligible. 
This justifies the use of the dimuon mass distribution to separate signal and background distributions of \ctau using the {\it sPlot\/}~\cite{CMS:2021puf} technique.
The \ctau resolution is obtained by fitting the negative tail of the \PJGy signal \ctau distribution in data using a sum of two Gaussian functions.
Note that \ctau can have negative values as a result of the finite detector resolution.
The shape of the \ctau distributions of prompt \PJGy signals is described by the \ctau resolution.
The \ctau distribution of nonprompt \PJGy signals is parameterized by the sum of three exponential functions convoluted with the resolution function.
The parameters of the three exponential functions are determined by a fit of the \ctau distribution of simulated nonprompt \PJGy mesons at the generator level.
The background \ctau distribution in data is fit to a sum of exponential functions.
To extract the yields of prompt and nonprompt \PJGy mesons, a fit to the $m_{\mumu}$ and \ctau distributions is performed, where the free parameters include the \PJGy meson yield, nonprompt \PJGy meson fraction, and background yield.
A more detailed description of the fitting procedure can be found in Ref.~\cite{CMS:2017uuv}.

For \PGyP{2S} mesons, the nonprompt component is reduced by placing tight constraints on the \ctau distribution.
The constraint, studied in simulated samples, rejects the nonprompt \PGyP{2S} contamination while maintaining a prompt \PGyP{2S} purity of at least 90\%.
The remaining contribution of nonprompt \PGyP{2S} mesons ranges from 5 to 10\% and is assigned as a systematic uncertainty for the nonprompt \PGyP{2S} contamination.

\section{Extraction of \texorpdfstring{{\vn}}{vn}}
\label{sec:v2ext}
The \vn ($n$ = 2 and 3) values of \PJGy and \PGyP{2S} candidates are determined using the scalar product (SP) method~\cite{Sirunyan:2021ay}.
The Q-vectors are defined in the complex plane $Q_{n} = \sum_{k=1}^{M} \omega_{k} \re^{in\phi_{k}}$, obtained using the tracker or calorimeters.
Here, $M$ is the multiplicity of particles in the tracker or the number of towers for HF;
$\phi$ is the azimuthal angle of the particle or the tower;
$\omega_{k}$ is the weighting factor, that is the \pt of a particle for the tracker or the transverse energy deposited in an HF tower. 
In this analysis, Q-vectors from three subevents are calculated using the tracker at the mid-pseudorapidity ($\abs{\eta} < 0.75$) and similarly the two HF calorimeters covering the forward ($3 < \eta < 5$, HF+) and backward ($-5 < \eta < -3$, HF$-$) regions.
Two different procedures are applied in extracting \vtwo of \PJGy and \PGyP{2S}. 
The procedure for extracting \PGyP{2S} \vtwo has been established in previous publications~\cite{Sirunyan:2021ay}, while a new method is applied to the extraction of \PJGy \vtwo with the advantage of avoiding making assumptions of background candidate \vtwo.
The \vn coefficient for \PJGy or \PGyP{2S} mesons is obtained as follows:
\begin{equation}
\label{eq:qsp}
{\vn}\left\{ {{\rm{SP}}} \right\} \equiv {\frac{\left\langle {{Q_{n}^{\PJGy,\PGyP{2S}}}Q_{nA}^*} \right\rangle }  {\sqrt {{\frac{\left\langle {Q_{nA}^{}Q_{nB}^*} \right\rangle \left\langle {Q_{nA}^{}Q_{nC}^*} \right\rangle } {\left\langle {Q_{nB}^{}Q_{nC}^*} \right\rangle }}} }} .
\end{equation}

The Q-vector of the \PJGy or \PGyP{2S} candidate is defined as $Q_{n}^{\PJGy} = e^{in\phi}$, where $\phi$ is the azimuthal angle of the candidate.
The subscripts $A$ and $B$ refer to either HF+ or HF$-$, depending on the rapidity of the \PJGy or \PGyP{2S} candidate.
Flattening and recentering procedures are applied to the Q-vectors relative to HF and the tracker for removing detector acceptance effects.
To avoid short-range non-collective correlations, the $\eta$ gap between the \PJGy or \PGyP{2S} candidate and the detector used for the subevent determination is required to be at least three units of rapidity~\cite{Sirunyan:2021ay,Sirunyan:2017plt,Luzum:Elimin}.
For this purpose, HF+ is selected for A (B) when \PJGy and \PGyP{2S} candidates are produced at negative (positive) rapidity.
The subscript $C$ denotes the subevents taken from the tracker.
The denominator in Eq.~(\ref{eq:qsp}) is the correction factor to remove the finite resolution effect of the detectors and finite final-state multiplicity.

To extract the prompt and nonprompt \PJGy meson \vn values, the \PJGy candidates are classified into fine \vn intervals of their flow vector SPs (i.e., Eq.~(\ref{eq:qsp}) but not averaged over all \PJGy candidates).
Then, a simultaneous invariant mass and \ctau fit, as described in Section~\ref{sec:SinalExt}, is performed in each \vn interval to extract the corresponding inclusive \PJGy meson yield and the nonprompt \PJGy meson fraction, and to obtain the SP distribution of both prompt and nonprompt \PJGy signal.
The prompt \PJGy meson \vn values can be obtained by:
\begin{equation}
\vn^{\text{prompt}} =\frac{\sum (1-f_i)\vni Y^i}{\sum (1-f_i)Y^i},
\end{equation}
while the nonprompt \PJGy meson \vn values can be obtained by:
\begin{equation}
\vn^{\text{nonprompt}} =\frac{\sum f_i\vni Y^i}{\sum f_iY^i}.
\end{equation}
Here \vni is the center of the $i$-th SP bin, while $f_i$ and $Y^i$ are the nonprompt \PJGy meson fraction and inclusive \PJGy meson yield in the same bin, respectively.

\begin{figure}[hbt!]
  \centering
    \includegraphics[width=0.48\textwidth]{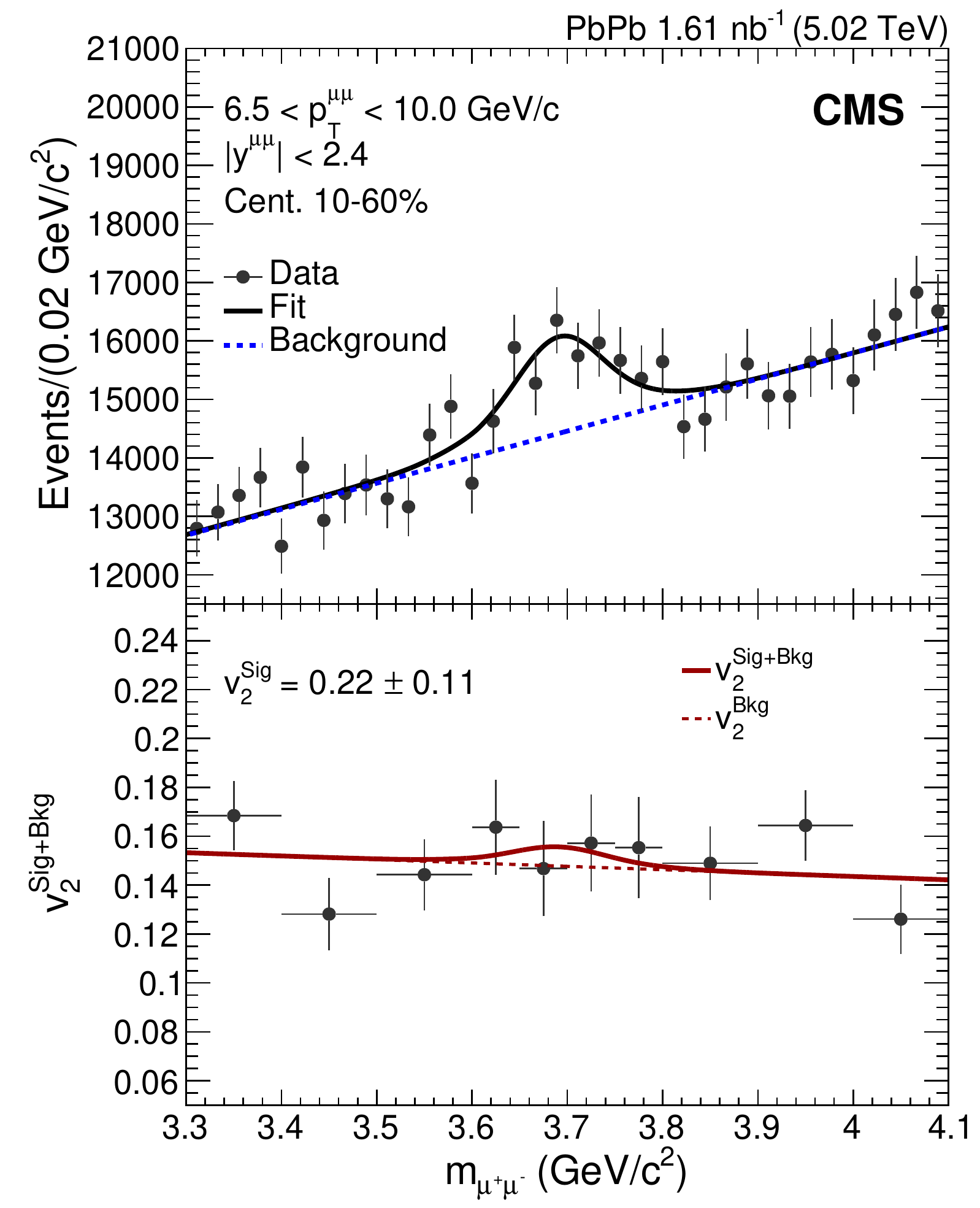}
    \includegraphics[width=0.48\textwidth]{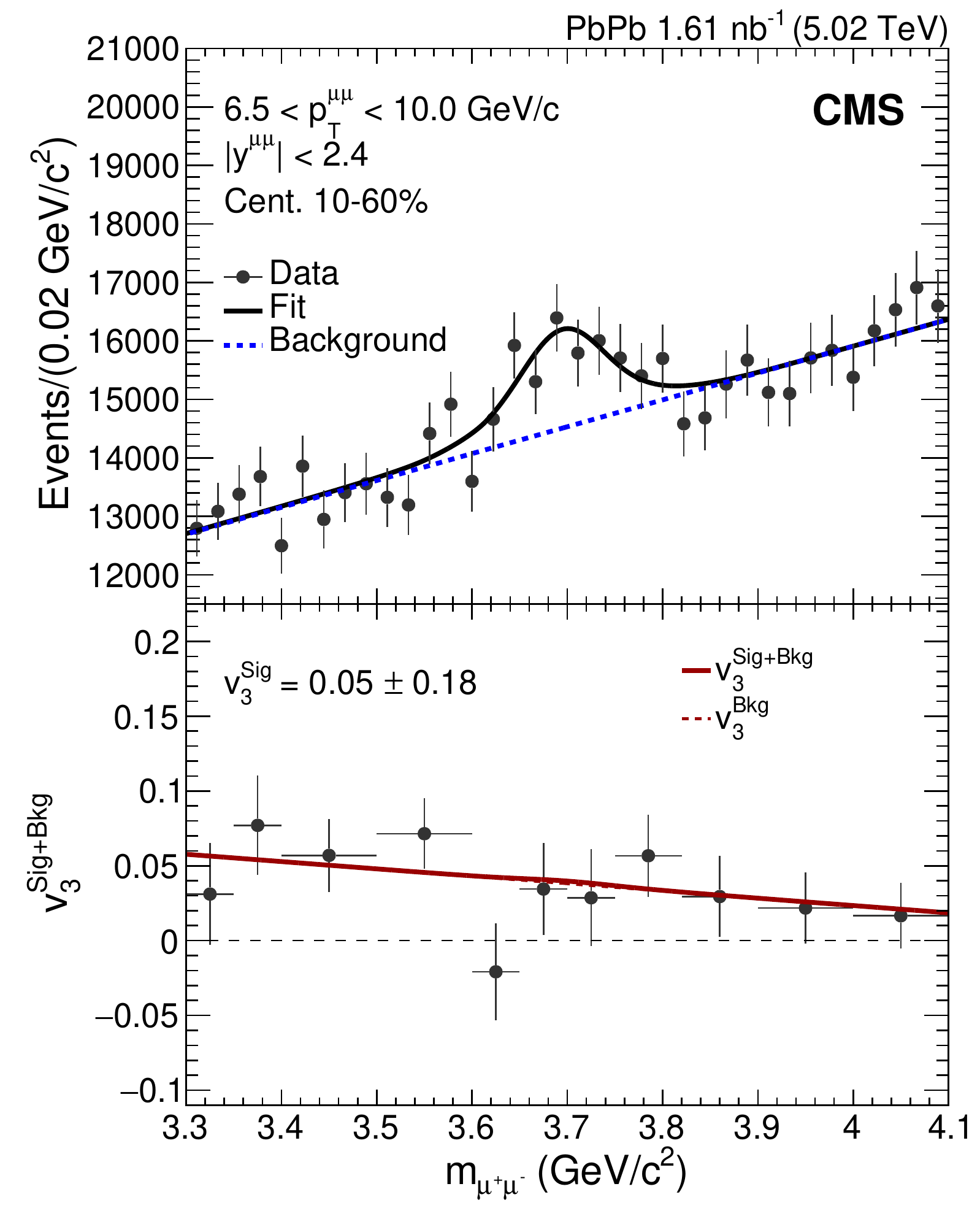}
	\caption{Simultaneous fits of the dimuon invariant mass spectrum and the $\vtwo^{\text{Sig+Bkg}}$ (\cmsLeft) and $\vthree^{\text{Sig+Bkg}}$ (\cmsRight) distributions for \PGyP{2S} mesons, as defined by Eq.~(\ref{vnvsmassequation}), for $6.5 < \pt < 10$ \GeVc and centrality 10--60\%. In the upper panel, the solid black (signal + background) and dashed blue (background only) lines show the result of the mass fit. In the lower panel, the vertical lines of the $v^{\text{Sig}}_{n}$ points represent the statistical uncertainty. The solid (\vn signal + background) and the dashed (\vn background) lines indicate the corresponding fit to the \vn distributions.}
    \label{fig:simulFits}
\end{figure}

A different method is used to extract the prompt \PGyP{2S} meson flow harmonics.
The average Q-vectors of all prompt \PGyP{2S} meson candidates are divided into bins of invariant mass, as shown in the lower panel of Fig.~\ref{fig:simulFits}.
For every invariant mass bin, the prompt \PGyP{2S} meson \vn coefficient is calculated using Eq.~(\ref{eq:qsp}).
The \PGyP{2S} candidate invariant mass spectrum and \vn distribution as a function of invariant mass are fit simultaneously using binned $\chi^{2}$ fits.
The invariant mass spectrum is fit with the sum of two CB functions and a Chebyshev polynomial function, as described in Section~\ref{sec:SinalExt}.
The \vn distribution is fit to the following formula:
\begin{equation}
\label{vnvsmassequation}
\vn^{\text{Sig+Bkg}}(m_{\text{inv}}) = \alpha(m_{\text{inv}})\vn^{\text{\PGyP{2S}}} + [ 1-\alpha(m_{\text{inv}}) ] \vn^{\text{Bkg}}(m_{\text{inv}}),
\end{equation}
where $\vn^{\PGyP{2S}}$ is the prompt \PGyP{2S} meson \vn, which is independent of $m_{\text{inv}}$ by definition; $\vn^{\text{Bkg}}(m_{\text{inv}})$ is the background \vn modeled as a first-order polynomial function of invariant mass; and $\alpha(m_{\text{inv}})$ is the \PGyP{2S} signal fraction as a function of invariant mass:
\begin{equation}
    \alpha(m_{\text{inv}}) = 
    \frac{ {\text{Sig}_{\text{\PGyP{2S}}}(m_{\text{inv}})} } { {[\text{Sig}_{\text{\PGyP{2S}}}(m_{\text{inv}}) + \text{Bkg}(m_{\text{inv}})]} }.
\label{vnvsmasseq2}
\end{equation}

An example of a simultaneous fit to the mass distribution (upper panel) and \vn profile (lower panel) for \PGyP{2S} mesons is shown in Fig.~\ref{fig:simulFits} for a representative analysis bin.

\section{Systematic uncertainties}
\label{sec:systematics}
Systematic uncertainties in this analysis originate from various sources including the modeling of signal and background pdfs, and acceptance and efficiency corrections (trigger, track reconstruction, and muon identification).
For uncertainties from the signal pdf, two factors are considered.
The first factor is the choice of the signal pdf, and the second is the values of fixed parameters in the invariant mass fit.
To evaluate uncertainty on the choice of two CB fit functions, an alternative pdf is formed by adding one CB function and one Gaussian function.
The deviation of \vn to the nominal one is quoted as the uncertainty.
Uncertainties from the fixed signal shape parameters are studied by releasing each parameter one at a time.
The root-mean-square (RMS) of differences between the nominal and altered \vn results is assigned as the uncertainty.
For the uncertainty in the background pdf, an exponential of a Chebyshev polynomial is considered as an alternative pdf for the invariant mass distribution.
The variation in the \vn results is quoted as the systematic uncertainty.

The uncertainty due to the acceptance correction is evaluated by comparing results with and without \pt reweighting of the signal shape in simulated samples.
The variation of efficiencies due to the T\&P correction is considered.
The change of results by varying the T\&P correction within its uncertainty is quoted as an uncertainty in the final \vn results.
In addition, the uncertainty due to hadronic event selection is considered.
The collision event filter is varied, thus allowing for a migration of \PJGy and \PGyP{2S} mesons across centrality boundaries.
The effect on measured \vn values is taken as a systematic uncertainty.
Then, the differences of resulting \vn values from the nominal ones are considered as the systematic uncertainty for the event selection.

Different components of the \ctau pdf are varied to estimate the uncertainty in the nonprompt \PJGy meson fraction.
The total (signal+background) distribution is used instead of individual distributions for the \ctau uncertainty distribution.
The \ctau resolution obtained from prompt \PJGy meson simulation is used instead of that evaluated from data. 
The parameters used in the function to describe the bottom hadron decay length shape of the nonprompt \PJGy meson simulations are released in the final fit to data.
Four exponential decay functions are also considered instead of three.
The maximum deviation between each pair of two variations is taken as an uncertainty for bottom hadron decay length distribution of the nonprompt \PJGy.
One additional exponential function on each side of the \ctau distribution of the background is used for the systematic uncertainty estimation.
The RMS of these four sources is taken as the systematic uncertainty for the nonprompt \PJGy fraction.

For the $v_{2}^{\text{Bkg}}(m_{\text{inv}})$ modeling of \PGyP{2S} mesons, a second-order polynomial is used as an alternative function.
The deviation of \vn results from nominal ones is assigned as the systematic uncertainty.
Uncertainties due to the nonprompt component in the \PGyP{2S} meson \vn are derived from the deviation of \vn by using a looser \ctau requirement that has twice the residual efficiency for nonprompt \PGyP{2S} mesons compared to the nominal case.
The total uncertainty, calculated for each bin of \pt or \Npart, is determined by taking the quadratic sum of each individual source of uncertainty.
In Table~\ref{tab:uncertainty}, the total systematic uncertainty is summarized along with its dominant sources.
The ranges provided are illustrative and represent the absolute values of the minimum and maximum uncertainties averaged over all \pt and \Npart bins.

\begin{table}[ht]
  \centering
  \topcaption{Summary of the absolute total systematic uncertainty and its dominant contributions, in units of $10^{-3}$.}
  \begin{tabular}{lcccccc} 
    \multirow{2}{*}{{Uncertainty source}} & \multicolumn{2}{c}{{Prompt \PJGy}} & \multicolumn{2}{c}{{Nonprompt \PJGy}} & \multicolumn{2}{c}{{Prompt \PGyP{2S}}}\\
                                          & {\vtwo} & {\vthree} & {\vtwo} & {\vthree} & {\vtwo} & {\vthree}\\ \hline
    {Total systematic}                    &{1.3--5.2}&{2.2--5.3}&{1.9--8.1 }&{3.3--9.1}&{32.0--63.0}&{30.0--82.0}\\
    {Signal parameters}                   &{0.4--3.0}&{0.1--3.0}&{0.9--5.0 }&{0.1--3.0}&{3.0--10.0}&{10.0--30.0}\\
    {Event selection}                     &{0.1--5.0}&{0.2--3.0}&{0.9--6.0 }&{0.3--2.0}&{10.0--30.0}&{5.0--20.0}\\
    {Nonprompt \PJGy fraction}            &{0.3--4.0}&{0.8--4.0}&{0.4--6.0 }&{2.0--9.0}&{\NA}&{\NA}\\
    {Nonprompt \PGyP{2S} contam.}         &{\NA}&{\NA}&{\NA}&{\NA}&{20.0--50.0} & {10.0--40.0}\\
    {Background \vn}                      &{\NA}&{\NA}&{\NA}&{\NA}&{1.0--20.0} & {0.7--50.0}\\
  \end{tabular}
  \label{tab:uncertainty}
\end{table}

\section{Results} 
\label{sec:results} 
\begin{figure}[hbt!p] 
\centering
    \includegraphics[width=0.99\textwidth]{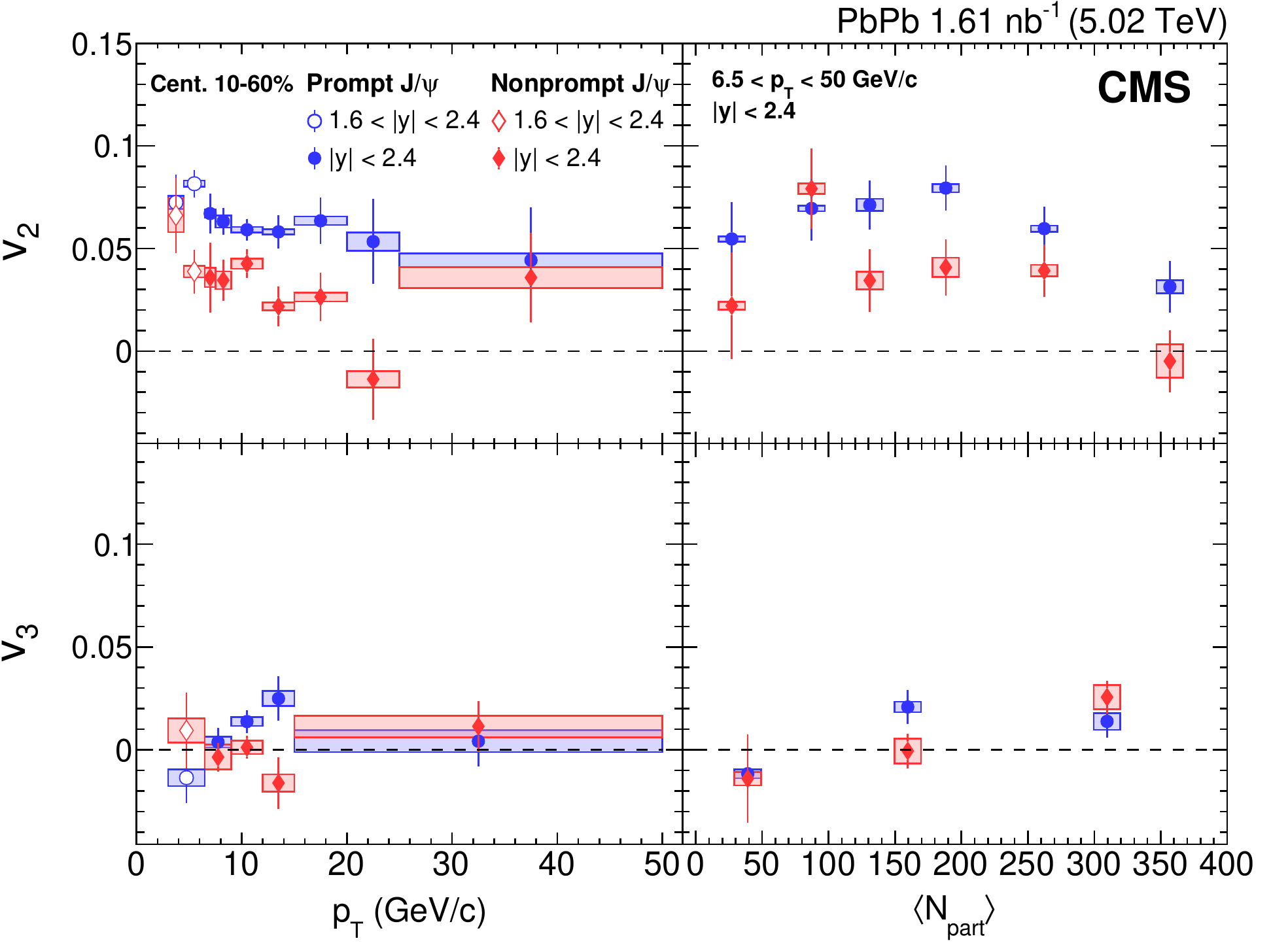}
    \caption{The \vtwo (upper) and \vthree (lower) values, as functions of \pt (\cmsLeft) and \Npart (\cmsRight) for prompt and nonprompt \PJGy mesons. 
The results for $3 < \pt < 6.5$ and $6.5 < \pt < 50$ \GeVc are studied in the rapidity range of $1.6 < \abs{y}< 2.4$ and $\abs{y} < 2.4$, respectively (\cmsLeft). 
The kinematic range for the \cmsRight panel is $6.5 < \pt < 50 \GeVc$ and $\abs{y} < 2.4$. 
The data points are positioned in the middle of each \pt bin. 
The vertical bars denote the statistical uncertainties and the rectangular bands show the systematic uncertainties.} 
    \label{fig:Jpsivnptcent} 
\end{figure}

The \vtwo and \vthree coefficients measured for prompt and nonprompt \PJGy mesons in \PbPb collisions are shown in Fig.~\ref{fig:Jpsivnptcent}, as functions of \pt and \Npart.
There is at most a weak \pt dependence observed for the prompt and nonprompt \PJGy mesons within their respective uncertainties.
Results in Fig.~\ref{fig:Jpsivnptcent} also show sizable \vtwo values for prompt \PJGy mesons even at high \pt, which are consistent with the \vtwo of charged hadrons in the high \pt region~\cite{CMS:2012tqw,Sirunyan:2018aa}.

The measured \vtwo values for nonprompt \PJGy mesons are smaller than those for prompt \PJGy mesons in the studied region.
Such a difference could not have been observed at $\sqrtsNN = 2.76 \TeV$~\cite{Khachatryan:2017sz} because of the limited amount of data.
These results are compatible with measurements from the ALICE and ATLAS Collaborations~\cite{ATLAS:Jpsiv2,Acharya:2020jf}, and extend to a higher \pt region.
A similar finding has been observed for prompt and nonprompt \PDz~\cite{CMS:2022vfn}.
These observations can be contrasted to those found for bottomonium measurements, in which the \vtwo values for \PgU mesons are found to be consistent with zero~\cite{Acharya:2019hlv,Sirunyan:2021ay}.

In Fig.~\ref{fig:Jpsivnptcent} (\cmsRight), the \vtwo values for prompt and nonprompt \PJGy mesons are found to be smallest in the most central \PbPb collision events (0--10\%), which corresponds to \Npart $>$ 356, and the \vtwo values increase toward mid-central collision events.
This is expected from the eccentricity in the initial collision geometry of these collisions.
The lower panel of the same figure shows measured \vthree values for prompt and nonprompt \PJGy components.
No significant nonzero \vthree values are found in studied kinematic intervals.

\begin{figure}[hbt!] 
\centering
    \includegraphics[width=0.68\textwidth]{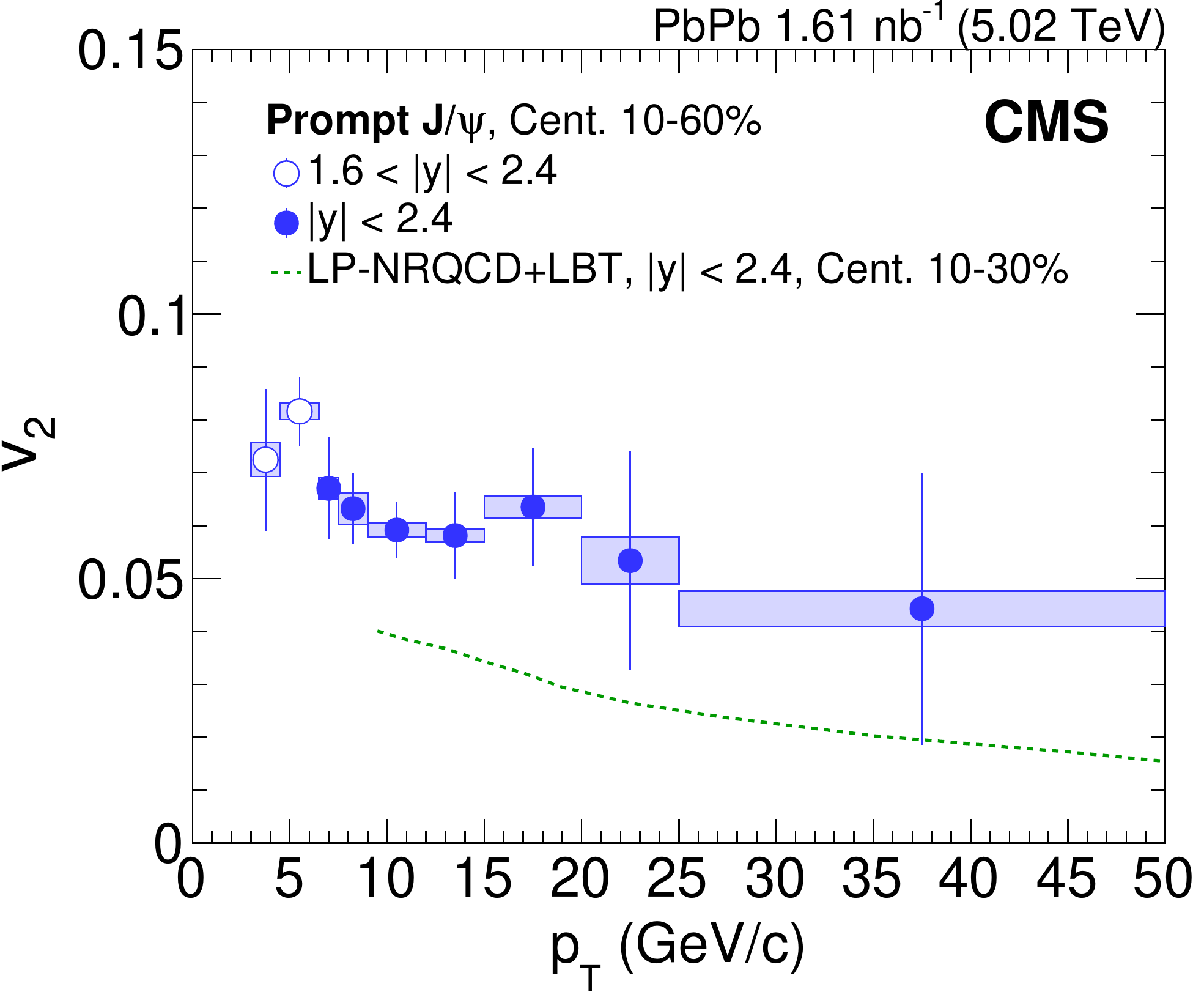}
    \caption{
The \vtwo values of prompt \PJGy mesons (blue circles) as a function of \pt in the 10--60\% centrality range compared with a model calculation in Ref.~\cite{Zhang:2022rby} (green dashed line). 
The results for $3 < \pt < 6.5$ and $6.5 < \pt < 50$ \GeVc are studied in the rapidity range of $1.6 < \abs{y}< 2.4$ and $\abs{y} < 2.4$, respectively. 
The vertical bars denote the statistical uncertainties and the rectangular bands show the systematic uncertainties.}
    \label{fig:JpsiTheory} 
\end{figure}

In Fig.~\ref{fig:JpsiTheory}, prompt \PJGy meson \vtwo values are compared to a theoretical prediction~\cite{Zhang:2022rby}, where high-\pt prompt \PJGy are produced from the jet fragmentation.
The medium response of jets in \PbPb collisions is simulated using the linear Boltzmann transport framework and the \vtwo values are calculated as a consequence of path length dependent energy loss of jets in the QGP.
Note that the model is calculated with the centrality interval 10--30\%, whereas the data is for the centrality interval 10--60\%.
A finite \vtwo value is predicted by the model calculation, suggesting that path length dependent energy loss of jets in this \pt range needs to be taken into account to describe the prompt \PJGy meson \vtwo.
However, the magnitude of \vtwo values in the data is found to be underpredicted by the model.
As no strong centrality dependence in mid-central collision events is observed in data, this difference between the observed \vtwo and theory calculation can not be explained by the different centrality intervals, indicating that additional contributions are needed to describe the prompt \PJGy meson \vtwo at high \pt.

\begin{figure}[hbt!] 
\centering
    \includegraphics[width=0.99\textwidth]{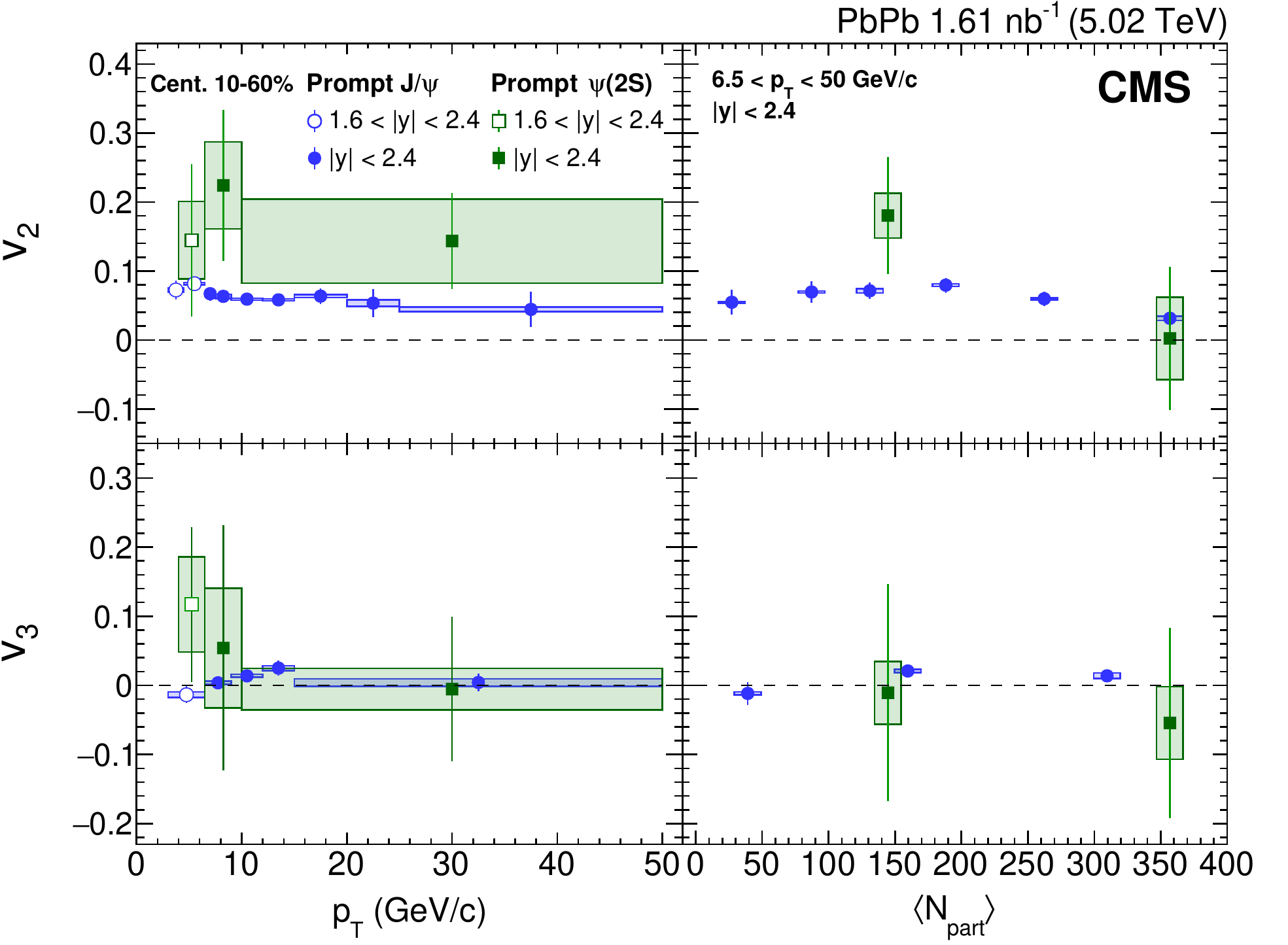}
    \caption{The \vtwo (upper) and \vthree (lower) values, as functions of \pt (\cmsLeft) and \Npart (\cmsRight) for prompt \PJGy (blue circles) and prompt \PGyP{2S} (green squares) mesons. 
The results for $\pt < 6.5$ and $\pt > 6.5$ \GeVc are studied in the rapidity range of $1.6 < \abs{y} < 2.4$ and $\abs{y} < 2.4$, respectively (\cmsLeft).
The kinematic range for the \cmsRight panel is $6.5 < \pt < 50 \GeVc$ and $\abs{y} < 2.4$.
The data points are positioned in the middle of each \pt bin.
The vertical bars denote the statistical uncertainties and the rectangular bands show the systematic uncertainties.}
    \label{fig:psi2svn} 
\end{figure}

Figure~\ref{fig:psi2svn} shows the first \vn measurements of the azimuthal anisotropy for prompt \PGyP{2S} mesons in heavy ion collisions together with the results for prompt \PJGy mesons.
The \vtwo values are found to be slightly larger for the prompt \PGyP{2S} than for the prompt \PJGy mesons, especially at higher \pt and in peripheral \PbPb collisions, but the statistical significance of this difference is too low to draw any conclusion, with p-values of at best 10\%.
The \vthree values are found to be consistent with zero in the measured bins.

\clearpage

\section{Summary}
\label{sec:conclusions}
The second-order (\vtwo) and third-order (\vthree) Fourier coefficients of azimuthal distributions for prompt and nonprompt \PJGy, and prompt \PGyP{2S} mesons are measured in \PbPb collisions as functions of transverse momentum (\pt) and event centrality.
The \vtwo values for prompt and nonprompt \PJGy mesons both indicate a decreasing trend from mid-central towards central collision events.
Within present experimental uncertainties, these \vtwo values show no clear \pt dependence between 3 and 50 \GeVc.
The prompt \PJGy meson \vtwo values are found to be larger than those of nonprompt \PJGy mesons throughout the studied kinematic region, suggesting different in-medium effects for charm and bottom quarks.
The prompt \PJGy meson \vtwo values are larger than those from a model calculation which considered the prompt \PJGy production above 10 \GeVc from jet fragmentation.
This discrepancy indicates additional effects are required to describe the observed sizable \vtwo at high \pt for prompt \PJGy mesons.
The \vthree values for prompt and nonprompt \PJGy mesons are reported for the first time, and found to be consistent with zero in the measured \pt and centrality intervals.
The \vtwo and \vthree values are also measured for prompt \PGyP{2S} mesons and are compatible with, although slightly larger than, those of prompt \PJGy mesons.
These \PJGy and \PGyP{2S} meson measurements provide new information about heavy quark interactions in the hot and dense medium created in heavy ion collisions.

\begin{acknowledgments}
We congratulate our colleagues in the CERN accelerator departments for the excellent performance of the LHC and thank the technical and administrative staffs at CERN and at other CMS institutes for their contributions to the success of the CMS effort. In addition, we gratefully acknowledge the computing centers and personnel of the Worldwide LHC Computing Grid and other centers for delivering so effectively the computing infrastructure essential to our analyses. Finally, we acknowledge the enduring support for the construction and operation of the LHC, the CMS detector, and the supporting computing infrastructure provided by the following funding agencies: BMBWF and FWF (Austria); FNRS and FWO (Belgium); CNPq, CAPES, FAPERJ, FAPERGS, and FAPESP (Brazil); MES and BNSF (Bulgaria); CERN; CAS, MoST, and NSFC (China); MINCIENCIAS (Colombia); MSES and CSF (Croatia); RIF (Cyprus); SENESCYT (Ecuador); MoER, ERC PUT and ERDF (Estonia); Academy of Finland, MEC, and HIP (Finland); CEA and CNRS/IN2P3 (France); BMBF, DFG, and HGF (Germany); GSRI (Greece); NKFIH (Hungary); DAE and DST (India); IPM (Iran); SFI (Ireland); INFN (Italy); MSIP and NRF (Republic of Korea); MES (Latvia); LAS (Lithuania); MOE and UM (Malaysia); BUAP, CINVESTAV, CONACYT, LNS, SEP, and UASLP-FAI (Mexico); MOS (Montenegro); MBIE (New Zealand); PAEC (Pakistan); MES and NSC (Poland); FCT (Portugal); MESTD (Serbia); MCIN/AEI and PCTI (Spain); MOSTR (Sri Lanka); Swiss Funding Agencies (Switzerland); MST (Taipei); MHESI and NSTDA (Thailand); TUBITAK and TENMAK (Turkey); NASU (Ukraine); STFC (United Kingdom); DOE and NSF (USA).
  
\hyphenation{Rachada-pisek} Individuals have received support from the Marie-Curie program and the European Research Council and Horizon 2020 Grant, contract Nos.\ 675440, 724704, 752730, 758316, 765710, 824093, 884104, and COST Action CA16108 (European Union); the Leventis Foundation; the Alfred P.\ Sloan Foundation; the Alexander von Humboldt Foundation; the Belgian Federal Science Policy Office; the Fonds pour la Formation \`a la Recherche dans l'Industrie et dans l'Agriculture (FRIA-Belgium); the Agentschap voor Innovatie door Wetenschap en Technologie (IWT-Belgium); the F.R.S.-FNRS and FWO (Belgium) under the ``Excellence of Science -- EOS" -- be.h project n.\ 30820817; the Beijing Municipal Science \& Technology Commission, No. Z191100007219010; the Ministry of Education, Youth and Sports (MEYS) of the Czech Republic; the Hellenic Foundation for Research and Innovation (HFRI), Project Number 2288 (Greece); the Deutsche Forschungsgemeinschaft (DFG), under Germany's Excellence Strategy -- EXC 2121 ``Quantum Universe" -- 390833306, and under project number 400140256 - GRK2497; the Hungarian Academy of Sciences, the New National Excellence Program - \'UNKP, the NKFIH research grants K 124845, K 124850, K 128713, K 128786, K 129058, K 131991, K 133046, K 138136, K 143460, K 143477, 2020-2.2.1-ED-2021-00181, and TKP2021-NKTA-64 (Hungary); the Council of Science and Industrial Research, India; the Latvian Council of Science; the Ministry of Education and Science, project no. 2022/WK/14, and the National Science Center, contracts Opus 2021/41/B/ST2/01369 and 2021/43/B/ST2/01552 (Poland); the Funda\c{c}\~ao para a Ci\^encia e a Tecnologia, grant CEECIND/01334/2018 (Portugal); the National Priorities Research Program by Qatar National Research Fund; MCIN/AEI/10.13039/501100011033, ERDF ``a way of making Europe", and the Programa Estatal de Fomento de la Investigaci{\'o}n Cient{\'i}fica y T{\'e}cnica de Excelencia Mar\'{\i}a de Maeztu, grant MDM-2017-0765 and Programa Severo Ochoa del Principado de Asturias (Spain); the Chulalongkorn Academic into Its 2nd Century Project Advancement Project, and the National Science, Research and Innovation Fund via the Program Management Unit for Human Resources \& Institutional Development, Research and Innovation, grant B05F650021 (Thailand); the Kavli Foundation; the Nvidia Corporation; the SuperMicro Corporation; the Welch Foundation, contract C-1845; and the Weston Havens Foundation (USA).
\end{acknowledgments}

\bibliography{auto_generated} 

\appendix
\section{Kinematic regions}\label{app:KinematicRegion}
The kinematic ranges for \PJGy and \PGyP{2S} mesons used in this study are summarized in Table~\ref{tab:KinematicRegion}.
The centrality and \Npart values used are calculated using a Glauber model simulation~\cite{LoizidesImprovedMC} and are summarized in Table~\ref{tab:CentralityNpart}.

\begin{table}[ht]
  \centering
  \topcaption{A Summary of the kinematic regions for \PJGy and \PGyP{2S} in the centrality interval 10--60\%.}
  \begin{tabular}{ccc}
                          & \PJGy                & \PGyP{2S}             \\ \hline
    $\abs{y} < 2.4$       & $6.5 < \pt < 50 \GeVc$ & $6.5 < \pt < 50 \GeVc$ \\
    $1.6 < \abs{y} < 2.4$ &  $3 < \pt < 6.5 \GeVc$ &  $4 < \pt < 6.5 \GeVc$ \\
  \end{tabular}
  \label{tab:KinematicRegion}
\end{table}

\begin{table}[ht]
  \centering
  \topcaption{The centrality classes and \Npart values for PbPb collisions at $\sqrtsNN = 5.02\TeV$ used in this study.}
  \begin{tabular}{cc}
    Centrality class & \Npart \\ \hline
    0--10\%          & 356.9 $\pm$ 0.9  \\
    10--20\%         & 262.3 $\pm$ 1.3 \\
    20--30\%         & 188.2 $\pm$ 1.4 \\
    30--40\%         & 131.0 $\pm$ 1.4 \\
    40--50\%         &  87.1 $\pm$ 1.3 \\
    50--90\%         &  27.1 $\pm$ 0.6 \\ [\cmsTabSkip]
    0--20\%          & 309.6 $\pm$ 1.0 \\
    20--40\%         & 159.6 $\pm$ 1.4 \\
    40--90\%         &  39.1 $\pm$ 0.7 \\ [\cmsTabSkip]
    10--60\%         & 144.6 $\pm$ 1.2
  \end{tabular}
  \label{tab:CentralityNpart}
\end{table}
\cleardoublepage \section{The CMS Collaboration \label{app:collab}}\begin{sloppypar}\hyphenpenalty=5000\widowpenalty=500\clubpenalty=5000
\cmsinstitute{Yerevan Physics Institute, Yerevan, Armenia}
{\tolerance=6000
A.~Tumasyan\cmsorcid{0009-0000-0684-6742}
\par}
\cmsinstitute{Institut f\"{u}r Hochenergiephysik, Vienna, Austria}
{\tolerance=6000
W.~Adam\cmsorcid{0000-0001-9099-4341}, J.W.~Andrejkovic, T.~Bergauer\cmsorcid{0000-0002-5786-0293}, S.~Chatterjee\cmsorcid{0000-0003-2660-0349}, K.~Damanakis\cmsorcid{0000-0001-5389-2872}, M.~Dragicevic\cmsorcid{0000-0003-1967-6783}, A.~Escalante~Del~Valle\cmsorcid{0000-0002-9702-6359}, P.S.~Hussain\cmsorcid{0000-0002-4825-5278}, M.~Jeitler\cmsAuthorMark{1}\cmsorcid{0000-0002-5141-9560}, N.~Krammer\cmsorcid{0000-0002-0548-0985}, L.~Lechner\cmsorcid{0000-0002-3065-1141}, D.~Liko\cmsorcid{0000-0002-3380-473X}, I.~Mikulec\cmsorcid{0000-0003-0385-2746}, P.~Paulitsch, F.M.~Pitters, J.~Schieck\cmsAuthorMark{1}\cmsorcid{0000-0002-1058-8093}, R.~Sch\"{o}fbeck\cmsorcid{0000-0002-2332-8784}, D.~Schwarz\cmsorcid{0000-0002-3821-7331}, M.~Sonawane\cmsorcid{0000-0003-0510-7010}, S.~Templ\cmsorcid{0000-0003-3137-5692}, W.~Waltenberger\cmsorcid{0000-0002-6215-7228}, C.-E.~Wulz\cmsAuthorMark{1}\cmsorcid{0000-0001-9226-5812}
% ...
\par}
\vskip\cmsinstskip
\dag:~Deceased\\
$^{1}$Also at TU Wien, Vienna, Austria\\
% ...
\end{sloppypar}
\end{document}